\documentclass[twocolumn,aps,prl,showpacs,floatfix,showkeys]{revtex4}

\usepackage{amsfonts}
\usepackage[english]{babel}
\usepackage{amsmath,amssymb}
\usepackage[all,cmtip]{xy}
\usepackage{pdfsync}
\def\beq{\begin{equation}}
\def\eeq{\end{equation}}
\def\bea{\begin{eqnarray}}
\def\eea{\end{eqnarray}}
\def\nn{\nonumber}
\newtheorem{proposition}{Proposition}
\newtheorem{theorem}{Theorem}

\newtheorem{remark}{\textbf{Remark}}

\newtheorem{defi}[theorem]{Definition}

\newcommand{\rd}{\mathrm{d}}
\begin{document}
\title[Bi--Hamiltonian manifolds and superintegrability]{Generalized Lenard Chains, Separation of Variables and Superintegrability}
\author{Piergiulio Tempesta}\email[E-mail: ]{p.tempesta@fis.ucm.es}
\affiliation{Departamento de F\'{\i}sica Te\'{o}rica II, Facultad de F\'{\i}sicas, Universidad
Complutense, 28040 -- Madrid, Spain}
\author{Giorgio Tondo}\email[E-mail: ]{tondo@units.it}
\affiliation{Dipartimento di Matematica e Geoscienze, Universit\`a  degli Studi di Trieste,
Piaz.le Europa 1, I--34127 Trieste, Italy}
%\msc{37J35}

\begin{abstract}

We show that the notion of generalized Lenard chains naturally allows  formulation of the theory of multi--separable and superintegrable systems in the context of bi--Hamiltonian geometry.  We prove that the existence of generalized Lenard chains generated by a Hamiltonian function defined on a four--dimensional $\omega N$ manifold guarantees the separation of variables. As an application, we construct such chains for the H\'enon--Heiles systems and for the classical Smorodinsky--Winternitz systems. New bi--Hamiltonian structures for the Kepler potential are found.
\end{abstract}

%\volumeyear{year}
%\volumenumber{number}
%\issuenumber{number}
%\eid{identifier}
\date{March 13, 2012}
%\startpage{1}
%\endpage{6}
\maketitle

\tableofcontents

\section{Introduction: superintegrability and separation of variables}

The aim of this paper is to establish a new connection between the theory of integrable and superintegrable systems on one side, and that of bi--Hamiltonian separation of variables on the other side. We will provide a theoretical framework for studying separation of variables for classical systems, by means of the notion of generalized Lenard (GL) chains. These chains, jointly with a couple of compatible Poisson tensors, are the main geometrical objects for our bi--Hamiltonian description of classical mechanics. These structures guarantee the separation of variables in a suitable bi--structured manifold.

In classical mechanics, superintegrable systems are Hamiltonian systems that possess more than $N$ integrals of motion functionally independent, globally defined in a $2N$--dimensional phase space (see e.g. \cite{TWR} for a monograph on the topic). These systems are also called noncommutatively integrable \cite{MF}, \cite{SV}. Especially important are the maximally superintegrable ones, i.e. those having $2N-1$ integrals. It turns out that for these systems all bounded orbits are closed and the motion is periodic \cite{Nekh}. The first study in this direction was made by Bertrand \cite{Bertrand}, who derived this result in the case of spherically symmetric potentials. Among the physically most relevant superintegrable potentials are the harmonic oscillator and the Kepler potential, the Calogero--Moser potential, the Smorodinsky--Winternitz systems, the Euler top, etc. \cite{TWR}--\cite{Evans}.

In quantum mechanics, superintegrable systems are also particularly interesting: they possess accidental degeneracy of the energy levels. This degeneracy can be removed by considering the quantum numbers associated with the additional integrals of motion. A paradigmatic example is offered by the Coulomb atom \cite{Fock}--\cite{JH}.

Recently, new examples of superintegrable systems have been discovered \cite{RTW1}--\cite{PostWint}. In \cite{Nekh}, \cite{Fasso}, it has been proved that the topology of phase space is characterized by a bifoliation consisting of an isotropic foliation of invariant tori, and of its coisotropic polar foliation.

One of the most effective methods to solve Hamiltonian systems is to find a complete integral of the corresponding Hamilton--Jacobi (HJ)  equation through the technique of separation of variables.  For the sake of clarity, we will recall briefly the geometric setting of Hamiltonian dynamics \cite{AM}.

Let $(M,\omega)$  be  a symplectic manifold, i.e.  a $2n$--dimensional  manifold endowed with a non degenerate
closed two--form $\omega$, said to be a symplectic form. Such a geometrical structure selects a privileged dynamics on $M$, the one given
by Hamiltonian vector fields defined by
\begin{equation*}
i_{X_H}\omega=-\rd H
\end{equation*}
($i_{X_H}$ denotes the contraction operator with regard to  the vector field  $X_H$ and $\rd$ denotes the exterior derivative operator)
 or, equivalently
\begin{equation*} \label{eq:Xfsymp}
X_H=(\omega^\flat)^{-1}\rd H  \ ,
\end{equation*}
 where $\omega^\flat:TM\rightarrow T^*M $ denotes
the fiber bundles isomorphism induced by $\omega$.  The function $H$ is said to be the Hamiltonian function of the vector field $X_H$.
A symplectic form acting on  vector fields is equivalent to a non degenerate Poisson  bracket defined as
\begin{equation} \label{eq:Pb}
\{F,G\}:=\omega(X_F,X_G)=<\rd F,X_G> \ ,
\end{equation}
($ < , >$ denotes the natural pairing between 1--forms and vector fields), i.e. as a skew--symmetric composition law on the ring $C^\infty(M)$
satisfying
\begin{eqnarray}
\nn\label{eq:PoisLeib}
\{F,GH\}&=&\{F,G\}H+ \{F,H\}G\hspace{90pt}\\
 \label{eq:PoisJacobi}
\nn 0&=&\{F,\{G,H\}\}+\{G,\{H,F\}\}\}+\{H,\{F,G\}\} \hspace{5pt}\\
\label{eq:Poisinv}
 \{F,G\}&=&0 \quad \forall F\implies \rd G=0.  \hspace{80pt}
\end{eqnarray}
 A local chart $(\mathbf{q},\mathbf{p}):=(q_1,\ldots,q_n,p_1,\ldots,p_n)$ satisfying
$\{q_i,p_j\}=\delta_{ij}$, $\{q_i,q_j\}=\{p_i,p_j\}=0$  is said to be a
system of canonical or Darboux coordinates. In such coordinates the
(time--independent) Hamilton--Jacobi equation corresponding to a Hamiltonian
vector field $X_H$  reads
\begin{equation} \label{eq:HJ}
H(q_1,\ldots,q_n,\frac{\partial W}{\partial q_1},
\ldots, \frac{\partial W}{\partial q_n})=E \ ,
\end{equation}

 A solution $(W,E)$ with $W(\mathbf{q}; a_1,\ldots,a_n)$,  $E(a_1, \ldots , a_n)$ and $\{a_i\}_{1\le i\le n}$  constant parameters,
 such that  $det[\frac{\partial^2 W}{\partial q_i \partial a_j}] \neq 0$, is said to be a
complete integral of Eq. (\ref{eq:HJ}) and allows one to solve \emph{locally} the equation of motions for $X_H$.  In fact,  a solution of Eq. (\ref{eq:HJ}) is the generating function of a canonical transformation that maps the Darboux coordinates $(\mathbf{q},\mathbf{p})$  into a new system of Darboux coordinates  $(\mathbf{\tilde q},\mathbf{\tilde p})$ satisfying the finite equation of motions
\begin{eqnarray} \label{eqqtilde}
{\tilde q^k}(t)&=&\frac{\partial E}{\partial a_k} t + {\tilde q^k}(0) \qquad\qquad k=1, \ldots , n  \\
\nn {\tilde p_k}&=&a_k
\end{eqnarray}
One can solve the following equation with respect to  $(\mathbf{q})$
\begin{equation*}
{\tilde q^k}=\frac{\partial W}{\partial a_k }(\mathbf{q}; a_1,\ldots,a_n),
\end{equation*}
and by taking into account Eq. (\ref{eqqtilde}), one can write down the finite equations of motion in the original coordinates. In addition, by solving with regard  to $\{a_i\}_{1\le i\le n}$ the following equations
  \begin{equation}
 p_k=\frac{\partial W}{\partial q_k }(\mathbf{q}; a_1,\ldots,a_n) \qquad\qquad k=1, \ldots , n
\end{equation}
the function $W$ provides   the $n$ involutive integrals  ${\tilde p_k}$,
 \begin{equation}
\label {eq:pInt}
 {\tilde p_k}=a_k=H_k(\mathbf{q},\mathbf{p}) \qquad\qquad  k=1,\ldots ,
\end{equation}
whose Hamilton--Jacobi equations admit, by construction, the same solution $(W,E)$ than Eq. \eqref{eq:HJ}.
If such integrals are globally defined, then $X_H$ is \textit{Liouville--integrable};
we assume this property throughout this work.
\par
Most of the cases in which a complete integral is explicitly found, occur when
$W$ is an additively separated function of the coordinates $q_i$
\begin{equation} \label{eq:sHJ}
W(\mathbf{q}; a_1,\ldots,a_n)=\sum_{i=1}^n W_i(q_i; a_1,\ldots,a_n) \ .
\end{equation}
In such a case $H$ is said to be separable and the coordinates
$(\mathbf{q},\mathbf{p})$ are said to be separated coordinates with regard  to $H$, in order to
stress that the possibility of finding a separated complete integral of Eq.
(\ref{eq:HJ}) depends on the choice of the coordinates.
\par
One of the classical issues in the  theory of
separation of variables (SoV) is to find criteria to decide
if a given  Hamiltonian function $H$ is separable in an assigned system of
canonical coordinates and, in the affirmative case, to find a separated complete
integral of the Hamilton--Jacobi equation.  In this regard, a prominent role is played by the test by
Levi--Civita  \cite{L}. It states that $H$ is separable in a Darboux chart $(\mathbf{q},\mathbf{p})$ if and only if the following $n(n-1)/2$ conditions, 

\begin{eqnarray}
\nonumber \noindent &&0=\frac{\partial H}{\partial q_i}\frac{\partial H}{\partial q_j}
\frac{\partial^2 H}{\partial p_i \partial p_j}+
\frac{\partial H}{\partial p_i}\frac{\partial H}{\partial p_j}
\frac{\partial^2 H}{\partial q_i \partial q_j}-
\frac{\partial H}{\partial q_i}\frac{\partial H}{\partial p_j}
\frac{\partial^2 H}{\partial p_i \partial q_j} \\
\noindent && -\frac{\partial H}{\partial q_j}\frac{\partial H}{\partial p_i}
\frac{\partial^2 H}{\partial p_j \partial q_i}, \qquad (1\le i< j\le n)
\end{eqnarray}

%\begin{equation} \label{eq:LCc}
%0=\frac{\partial H}{\partial q_i}\frac{\partial H}{\partial q_j}
%\frac{\partial^2 H}{\partial p_i \partial p_j}+
%\frac{\partial H}{\partial p_i}\frac{\partial H}{\partial p_j}
%\frac{\partial^2 H}{\partial q_i \partial q_j}-
%\frac{\partial H}{\partial q_i}\frac{\partial H}{\partial p_j}
%\frac{\partial^2 H}{\partial p_i \partial q_j}-
%\frac{\partial H}{\partial q_j}\frac{\partial H}{\partial p_i}
%\frac{\partial^2 H}{\partial p_j \partial q_i}
%\end{equation}
are satisfied.
\par
As we wish to study separable Hamiltonian systems  that are Liouville--integrable,  in principle we can start
 with a set of $n$ independent Hamiltonian functions in involution with regard to  the Poisson brackets
(\ref{eq:Pb}). In this framework, in the
tradition of the {\it Italian school},  an important result has been obtained by Benenti  in \cite{Ben80}. It gives a characterization of separated coordinates
in terms of Poisson bracket.
%\begin{theorem} {\rm (Benenti, 1980)} \label{th:Ben}

The Hamiltonian functions $\{H_i\}_{1\le i\le n}$ are separable in a set of
canonical coordinates $(\boldsymbol{q},\boldsymbol{p})$ if and only if they
are in {\rm separable } involution, i.e. if and only if they satisfy
\begin{equation} \label{eq:SI}
\{H_i,H_j\}_{\vert k}=\frac{\partial H_i}{\partial q_k}
\frac{\partial H_j}{\partial p_k}-\frac{\partial H_i}{\partial p_k}
\frac{\partial H_j}{\partial q_k}=0 \ , \quad 1\le k \le n,
\end{equation}
where no summation over $k$ is understood.
%\end{theorem}
 However, such a theorem as well as the Levi--Civita test are not constructive,  since they do not help to find a complete integral of the
Hamilton--Jacobi equation (\ref{eq:HJ}).
In contrast, a constructive definition of SoV was given by Sklyanin  \cite{Skl} within the framework of Lax systems.
\begin{defi} \label{defi:Sk}
The Hamiltonian functions $\{H_i\}_{1\le i\le n}$ are separable in a set of
canonical coordinates $(\boldsymbol{q},\boldsymbol{p})$ if there exist
$n$ equations
\begin{equation} \label{eq:Sk}
\Phi_i(q_i,p_i; H_1,\ldots,H_n)=0 \qquad i=1,\ldots,n
\end{equation}
such that ${\rm  det}[\frac{\partial \Phi_i}{\partial H_j }] \neq 0$.
They are said to be  Sklyanin separation equations for $\{H_i\}_{1\le i\le
n}$ and allow one to construct a solution $(W,E)$ of the HJ equation (\ref{eq:HJ}). In fact, solving (\ref {eq:Sk})
with regard to  $p_k=\frac{\partial W_k}{\partial q_k }$, we get
\begin{equation}
W= \sum \int {p_k(q' _k; H_1, \ldots, H_n)_{\mid _{H_i=a_i} }dq'_k}
\end{equation}
\end{defi}

However,  the three above--mentioned criteria  of separability are not intrinsic
since they require   explicit knowledge of  the coordinates $(\boldsymbol{q},\boldsymbol{p})$ in order to be applied.
Recently, a new geometric approach to SoV has been developed, based on the bi--Hamiltonian
theory (\cite{MT}, \cite{FalquiPedroni}) and on GL chains (\cite{FMT}--\cite{Tondo}). It has succeeded in giving
intrinsic and constructive criteria of separability and has connected the classical
theory of SoV with the modern theory by Sklyanin. The bi--Hamiltonian  theory of SoV
is formulated in phase spaces represented by manifolds endowed with two geometric structures satisfying two suitable compatibility
conditions. Such structures are a symplectic form $\omega$ and a Nijenhuis (or hereditary) operator $N$ acting on the tangent bundle
of $M$. For this reason such manifolds have been  called $\omega N$ manifolds.
Whilst the symplectic form defines the algebra of Hamiltonian vector fields, the Nijenhuis operator defines sets of distinguished coordinates  that are separated coordinates for a special class of Hamiltonian vector fields, those belonging to GL chains, so called as  they are extensions  of classical Lenard  chains, widely known in soliton literature \cite{MagriLE}, \cite{PS}.

The main result of this paper is the following: we prove that, given a generic integrable system on the cotangent bundle of the Euclidean plane, the existence of a GL chain ensures separation of variables on  a $\omega N$ manifold.
We will study explicitly examples of many important physical systems, like the H\'enon--Heiles integrable models, the Smorodinsky--Winternitz systems and the Kepler potential. For all these systems we will construct explicitly bi--Hamiltonian structures. At the best of our knowledge, some of the structures we obtain are new.

Other studies concerning the bi--hamiltonian geometry of specific integrable systems that we study in this paper, are available in the literature \cite{CRG}--\cite{Ranada}.  A detailed comparison between our results and other approaches is performed punctually in the subsequent discussions of the physical examples. One of the main features of our approach is that it provides at once compatible Poisson structures and separated coordinates.

The paper is organized as follows. In Section 2, the theory of bi--Hamiltonian manifolds is reviewed. In Section 3, the main geometrical object of our theory, i.e. the GL chains, are introduced. A general theorem on the separation of variables for systems in $T^*E_2$ is proposed. In Section 4, the previous theory is applied to the study of general classes of integrable systems in the Euclidean plane. In Section 5, the bi--Hamiltonian geometry of the integrable H\'enon--Heiles systems is obtained. In Section 6,  GL chains are constructed for  the classical Smorodinsky--Winternitz systems and in particular for the Kepler potential. Some conclusions are drawn in the final Sec. 7.

\section{Bi--Hamiltonian manifolds and $\omega N$ manifolds}

 Generally, a Poisson bracket (see, e.g., \cite{Vaisman}) is defined as a skew--symmetric composition law  on
$C^\infty(M)$ which satisfies only the first two equations of  the system \eqref{eq:Poisinv} and not necessarily the third one. Equivalently, it can be defined by a
Poisson  bi--vector field, i.e.,  a skew--symmetric linear map
$P:T^*M\mapsto TM$ such that
\begin{equation*}
\label{ }
\{F,G\}_P:=<\rd F, P \rd G> \ , 
\end{equation*}
with a vanishing Schouten bracket
\begin{equation} \label{eq:PoisSch}
0=\left[ P, P \right ]_S(\alpha,\beta) := \mathcal{L}_{P\beta}(P)\alpha+
P (i_{P\alpha}\rd \beta) \quad  \alpha,\beta \in T^*M,
\end{equation}
% by means of
%\begin{equation}
%\{f,g\}:=<\rd f,P \rd g>
%\end{equation}
($\mathcal{L}$ denotes the Lie derivative). In the special case of symplectic manifolds,
$P:=(\omega^\flat)^{-1}$ is a Poisson bi-vector.
Generalizing  (\ref{eq:Pb}) the vector field $X_G:=P\, \rd G$ is said to be the
Hamiltonian vector field with Hamiltonian function $G$.
% It is an immediate  consequence of  (\ref{eq:PoisSch}) that each Hamiltonian vector field is a symmetry of $P$
%\begin{equation}\label{eq:Psym}
%\mathcal{L}_{X_g}P=0 \qquad\qquad \forall g \in C^\infty(M) \ .
%\end{equation}
\par
Bi--Hamiltonian manifolds were introduced by Magri \cite{Magri78} as models of phase
space for  soliton equations.
\begin{defi}\label{def:biH}
A bi--Hamiltonian manifold $(M,P_0,P_1)$ is a manifold $M$ endowed with
 {\em two} Poisson bi-vectors fields such that
\begin{eqnarray} \label{eq:PQSch}
\nonumber \noindent &&0=2\left[ P_0, P_1 \right ]_S (\alpha , \beta):= \mathcal{L}_{P_0\beta}(P_1)\alpha+
P_1 (i_{P_0\alpha}\rd \beta)+\\
\noindent &&\mathcal{L}_{P_1\beta}(P_0)\alpha+
P_0 (i_{P_1\alpha}\rd \beta).
\end{eqnarray}
\end{defi}
Such a condition assures that the linear combination
$P_1-\lambda P_0$ is a  Poisson {\em pencil}, i.e. it is a Poisson bi-vector for
each
$\lambda
\in
\mathbb{C}$, and therefore the corresponding bracket $\{ ,\}_{P_1}-\lambda \{,
\}_{P_0}$ is a pencil of Poisson brackets. Condition (\ref{eq:PQSch}) is known as the
compatibility condition between $P_0$ and $P_1$.
\par
What happens if  one of the Poisson tensors, say $P_0$, is invertible,
and therefore its inverse is a symplectic operator $\omega^\flat:=P_0^{-1}$?
 In this case,  the bi--Hamiltonian manifold $M$ turns out to be an
$\omega N$ manifold (see \cite{MM}). Indeed,
the composed operator
$N:=P_1P_0^{-1}$,  thanks to the compatibility condition between $P_0$ and $P_1$, is a Nijenhuis (or
hereditary) operator compatible with the symplectic form $\omega$.
\begin{defi} \label{def:oN}
 A $\omega N$ manifold $(M, \omega, N)$ is a symplectic manifold endowed with   an endomorphism of the tangent bundle of $M$,
  $N:TM\mapsto TM$ which satisfies the following  conditions:
\begin{itemize}
\item
its  Nijenhuis torsion vanishes identically, i.e. $\forall X,Y \in TM$
\begin{equation} \label{eq:Ntorsion}
[NX,NY]-N([X,NY]+[NX,Y])+N^2[X,Y])=0;
\end{equation}

\item
it is compatible with $\omega$, i.e. the tensor $P_1=N (\omega ^\flat)^{-1}$ is again a Poisson tensor and is compatible with
$P_0:=(\omega ^\flat)^{-1} $, according to  Definition \ref{def:biH}.
\end{itemize}
 \end{defi}
In short, the condition \eqref{eq:Ntorsion} can be rephrased by saying that the endomorphism $N$ is a  Nijenhuis (or hereditary or recursion) operator.
\noindent
The adjoint  linear map with regard to  the natural pairing   will be denoted by $N^{T}: T^*M \mapsto T^*M$ and will be defined by
$$<N^T \alpha , X>=<\alpha ,NX>.$$

The condition  (\ref{eq:Ntorsion}) on the  endomorphism $N$,  introduced by Nijenhuis \cite{Nij}, has a relevant geometrical meaning: it implies  that the distributions of its eigenvectors  are integrable according to the Frobenius theorem.  Consequently, under suitable completeness assumption to be introduced below, one can  select  local coordinate charts,  half of the coordinates  being just the eigenvalues of $N$,  in  which $N$ takes a  diagonal form.
We suppose that at each point $x$  (or in a
dense open subset) of $M$, the Nijenhuis tensor field $N$ admits
$n$ distinct eigenvalues $\lambda_i (x) \ (i=1. \ldots, n)$ (maximally distinct). Since in a generic $\omega N$ manifold the
 eigenspaces of $N$ are even--dimensional, belonging to the kernel of the skewsimmetric tensor field  $P_1 -\lambda P_0$, from the above assumption  it follows that
 $N$ (and the adjoint tensor $N^T$) can be put in diagonal form. Also, the eigenvalues  $\lambda_i (x)$ can be chosen  as coordinate functions in a neighborhood of $x$, if we assume that they are functionally independent in $x$, i.e.
\begin{equation}
\rd\lambda_1(x)\wedge  \ldots   \wedge \rd\lambda_n(x) \neq 0 \ .
\end{equation}

%\begin{defi} \label{defi:rpoint}
Let $x$ be a point of an $\omega N$ manifold. It will be  called a \textit{regular point} if
the eigenvalues of $N$ are maximally distinct and functionally independent in $x$.
\par
From the semicontinuity of the rank function it follows that  there exist a suitable neighborhood of $x$ whose points are regular points.
%\end{defi}
Then we have the following result, proved in \cite{Turiel},  \cite{MMars}.
%\begin{proposition} { \rm (Magri \& Marsico, Turiel) } \label{prop:DNc}
 Let $(M,\omega , N)$ be an $\omega N$ manifold. In a suitable open neighborhood of a
regular point, the $n$ functions $\lambda_i (x)$ can be completed by {\rm quadratures}
with $n$ functions
$\mu_i (x)$  such that the chart
$(\boldsymbol{\lambda},\boldsymbol{\mu})$ is a Darboux chart for $\omega$
and,  moreover,
\begin{equation} \label{eq:Dprop}
 N^T \rd \lambda_i=\lambda_i\rd
\lambda_i \qquad \qquad
N^T \rd \mu_i=\lambda_i\rd \mu_i \ .
\end{equation}
%\end{proposition}
Then, in a Darboux chart $(\boldsymbol{\lambda},\boldsymbol{\mu})$
the Nijenhuis tensor $N$ takes a diagonal form
%\begin{equation} \label{eq:NDN}
%N=
%\begin{bmatrix}
%\boldsymbol{\Lambda}_n&\boldsymbol{0}_n\\
%\boldsymbol{0}_n&\boldsymbol{\Lambda}_n
%\end{bmatrix} \ ,
%\end{equation}
and  the coordinates
$(\boldsymbol{\lambda}, \boldsymbol{\mu})$ are said to be Darboux--Nijenhuis
(DN) coordinates. The Poisson tensor $P_1:=N
(\omega^\flat)^{-1}$ in DN coordinates takes the form
\begin{equation} \label{eq:P1DN}
P_1=
\begin{bmatrix}
\boldsymbol{0}_n&\boldsymbol{\Lambda}_n\\
-\boldsymbol{\Lambda}_n&\boldsymbol{0}_n
\end{bmatrix} \ ,
\end{equation}
where $\boldsymbol{\Lambda}_n={\rm diag}(\lambda_1,\cdots,\lambda_n)$ and
$\boldsymbol{0}_n$ is the $n\times n$ matrix with zero entries. The Darboux--Nijenhuis coordinates are just separation coordinates in the bi--Hamiltonian theory of SoV. Hereafter, with an abuse of notation, we will identify an operator $P$ with its matrix in a suitable basis.

\begin{remark} \label{rem:st}
It can be shown that the \emph{separated} canonical transformations
\begin{equation}
\label{eq:DNt}
q_i=f_i(\lambda_i) \ , \quad p_i=\frac{\mu_i}{f'_i} \ ,
\end{equation}
with $f_i$ a generic invertible smooth function of a single coordinate $\lambda_i$
preserve the property \eqref{eq:Dprop}, i.e.,

\begin{equation} \label{eq:DTprop}
 N^T \rd q_i=\lambda_i\rd q_i \qquad \qquad
N^T \rd p_i=\lambda_i\rd p_i \ .
\end{equation}
From a geometrical point of view, we can say that coordinates
$(\boldsymbol{\lambda},\boldsymbol{\mu})$ and
$(\boldsymbol{q},\boldsymbol{p})$, related by transformations \eqref{eq:DNt} are adapted to the same coordinate web \cite{Bweb}.
\end{remark}
\section{Generalized Lenard chains}

After having introduced the geometrical structures which define separation coordinates in the bi--Hamiltonian theory of SoV, let us characterize the class of Hamiltonian functions which are separable in DN coordinates. For the sake of concreteness, we will do this  in the case of a four--dimensional manifold, warning that it has been generalized to a $n$ dimensional manifold \cite{FMT}, \cite{FalquiPedroni}, \cite{Tondo}.

\begin{theorem} \label{th:maint}
Let  $(M, \omega,  N)$ be  a four--dimensional   $ \omega N$  manifold and $(\lambda_1, \lambda_2, \mu_1, \mu_2)$  a DN local chart.
Let  $H$ be  a smooth function in $M$. The DN coordinates $(\lambda_1, \lambda_2, \mu_1, \mu_2)$ are separated variables for $H$  if and only if there exist two smooth  functions $f$ and  $g$ such that the one form

\begin{equation} \label{eq:GLC}
\alpha= f \, \rd H+ g\,  N^T \rd H
\end{equation}
is an exact one form, i.e., $\alpha$ is the differential of a function, say $H_2$
\begin{equation} \label{cc}
\alpha=\rd H_2.
\end{equation}
In this case, the function $H_2$ is an integral of motion in involution with $H_1:=H$ and the same DN coordinates are separated variables for $H_2$ as well.
%Moreover, $H_1$ and $H_2$ are in bi--involution, i.e. are in involution w.r.t. both $\{ ,\}_{P_0}$ and $\{,\}_{P_1}$.%
\end{theorem}
\textbf{Proof}. In the above--mentioned chart, $N$ takes the diagonal form

\begin{equation}
\label{eq:2gN}
N= \lambda_1(\frac{\partial}{\partial \lambda_1}\otimes \rd \lambda_1+\frac{\partial}{\partial \mu_1}\otimes \rd \mu_1)+\lambda_2(\frac{\partial}{\partial \lambda_2}\otimes \rd \lambda_2+\frac{\partial}{\partial \mu_2}\otimes \rd \mu_2)
\end{equation}
%\begin{equation} \label{eq:2gN}
%N=diag(\lambda_1, \lambda_2, \lambda_1, \lambda_2)
%\end{equation}
Let us suppose that Eqs. (\ref{eq:GLC}) and (\ref{cc}) are fulfilled.  Then, it follows that
\begin{equation}\label{eq:GLCcoord}
\begin{array}{lll}
  \frac{\partial H_2}{ \partial \lambda_k}&=&f  \frac{\partial H_1}{ \partial \lambda_k}+g\,  \lambda_k \frac{\partial H_1}{ \partial  \lambda_k} \\
\frac{\partial H_2}{ \partial \mu_k}&=&f  \frac{\partial H_1}{ \partial  \mu_k}+g\,  \lambda_k \frac{\partial H_1}{ \partial  \mu_k}
\end{array}
\end{equation}
with $k=1,2$. Therefore
\begin{equation} \label{eq:Ben4}
\{H_1, H_2\}_{|k}= \frac{\partial H_1}{ \partial  \lambda_k}\frac{\partial H_2}{ \partial  \mu_k}-
\frac{\partial H_1}{ \partial  \mu_k} \frac{\partial H_2}{ \partial  \lambda_k} \stackrel{(\ref{eq:GLCcoord})}{=}0 \ ,
\end{equation}
i.e.,  $H_1$ and $H_2$ are in separable involution according to Benenti's theorem (see formulas (\ref{eq:SI})), w.r.t.   a DN chart. Therefore, $H_2$ is an integral of motion for $X_H$.
\par
Viceversa, let us suppose that $(\lambda_1, \lambda_2, \mu_1, \mu_2)$ are separated variables for $H_1$ and $H_2$ and let us consider the equation
\begin{equation}\label {ccc}
f \rd H_1+g N^T \rd H_1= \rd H_2
\end{equation}
in the unknown functions $f$ and $g$. In the local chart $(\lambda_1, \lambda_2, \mu_1, \mu_2)$, Eq. (\ref{ccc}) takes the form
\begin{eqnarray}
\label{sys1}
f +g \lambda_1&=&\frac{\frac{\partial H_2}{\partial  \lambda_1}}{\frac{\partial H_1}{\partial \lambda_1}}, \qquad
f +g \lambda_2 =\frac{\frac{\partial H_2}{\partial  \lambda_2}} {\frac{\partial H_1}{\partial  \lambda_2}}, \\
\label{sys2}
 f +g \lambda_1&=&\frac{\frac{\partial H_2}{\partial \mu_1}} {\frac{\partial H_1}{\partial  \mu_1}}, \qquad
f +g \lambda_2=\frac{\frac{\partial H_2}{\partial  \mu_2}} {\frac{\partial H_1}{\partial  \mu_2}} \ .
\end{eqnarray}
We observe that   the first equations (\ref{sys1}) and (\ref{sys2}) coincide, so do the second equations (\ref{sys1}) and (\ref{sys2}), thanks to the  conditions  (\ref{eq:Ben4}). Thus the above system of four equations  reduces to two equations  that admit the unique solution
\begin{eqnarray}
f&=&\frac{1}{ \lambda_2- \lambda_1}(\lambda_2 \frac{\frac{\partial H_2}{\partial  \lambda_1}} {\frac{\partial H_1}{\partial  \lambda_1}} -
 \lambda_1 \frac{\frac{\partial H_2}{\partial  \lambda_2}} {\frac{\partial H_1}{\partial  \lambda_2}}),\\
g&=&\frac{1}{ \lambda_2- \lambda_1}(- \frac{\frac{\partial H_2}{\partial \lambda_1}} {\frac{\partial H_1}{\partial \lambda_1}}
+\frac{\frac{\partial H_2}{\partial  \lambda_2}} {\frac{\partial H_1}{\partial  \lambda_2}} ).
\end{eqnarray}

%\end{proof}
%\begin{defi}
Then we will say that Hamiltonian functions related by Eqs. (\ref{eq:GLC}) and (\ref{cc})  belong to a GL chain generated by $(\omega, N, H)$
since, for $(f=0, g=1)$, a GL chain reduces to a classical Lenard chain.
\begin{remark}
We observe that, if $\left( f=-(\lambda_1+ \lambda_2), g=1 \right)$, we get a quasi--bi--Hamiltonian (QBH) chain of Pfaffian type \cite{BCRR, MT} generated by the function $H$.
%We observe that, if $\left(f=0, g=1/(\lambda_1 \lambda_2)\right)$, we get a Quasi--Bi--Hamiltonian (QBH) %chain of Pfaffian type \cite{BCRR, MT} generated by the function $H_2$.
\end{remark}
%\end{defi}
\par
Theorem \ref{th:maint}
suggests the following procedure in order to classify Hamiltonian systems separable in DN coordinates:
\begin{enumerate}
  \item
  Choose a Darboux  chart $(q_1, q_2, p_1,p_2)$ in a $4$--dimensional symplectic manifold $M$.
  \item
  Construct a $\omega N$ structure which has $(q_1, q_2, p_1,p_2)$ as DN coordinates.
  \item
  Search for Hamiltonian function $H$ and for functions $f$ and $g$ such that make the one form (\ref{eq:GLC}) exact.
\end{enumerate}

This procedure can be considered as an \textit{inverse problem}, with respect to the \textit{direct approach} that starts from a given Hamiltonian and aims to find separation coordinates.

Let us observe that the above method provides the integral of motion $H_2$ together with a set of separated variables both for $H$ and $H_2$.

\section{Bi--Hamiltonian geometry in $T^*E_{2}$: construction of GL chains}

In this section, we wish to apply the procedure previously discussed to the study of the bi--Hamiltonian properties of systems defined in the cotangent bundle of the Euclidean plane.  Although we recover well--known results, the motivation to study this preliminary case is that it allows us to derive the bi--Hamiltonian structures for systems separating in one coordinate system in a transparent way. The results  derived here will be used in the following sections to classify multi--separable systems and to derive their geometrical properties. Precisely, we will study natural Hamiltonian functions
$$
H=\text {kinetic energy }+ \text {potential energy}
$$
and we will recover the most general form of the potential that makes $H$ separable in Cartesian, polar and parabolic coordinates in the Euclidean plane $E_{2}$. Here
 $M=T^*E_{2}$.

\subsection{Classical Separation of Variables}
\subsubsection{Cartesian case}
Let us consider the natural Hamiltonian function
\begin{equation}
H=\frac{1}{2} \left(p_x^2+p_y^2\right)+V(x,y),
\end{equation}
$(x, y,p_x, p_y) $ being Cartesian coordinates and conjugate momenta.
According to the requirement \eqref{eq:2gN}, we choose  the Nijenhuis tensor $N_{car}: T(T^*E_2) \rightarrow  T(T^*E_2)$
\begin{equation*}\label{matice}
N_{car}=\mathrm{diag}(x,y,x,y).
\end{equation*}
The one--form  (\ref{eq:GLC}) reads
\begin{equation}
\label{eq:alphacart}
\alpha=f\left(x,y,p_x,p_y\right)dH+g\left(x,y,p_x,p_y\right)N^{T}dH,
\end{equation}
and the closure condition $d\alpha =0$ provides a system of nonlinear partial differential equations (PDEs) for $f$, $g$ and $V$ reported in the Appendix, formula (\ref{cartA1}). By combining Eqs. (\ref{cartA1}), we deduce the interesting differential consequence:
\begin{equation} \label{Vcart}
\left(y-x\right)gV_{xy}=0.
\end{equation}
The case $g=0$ is trivial, since it would imply dependence of $\alpha$ only on $dH$. So, if $g\neq 0$ and $x \neq y$,   Eq. (\ref{Vcart}) implies  $V_{xy}=0$, i.e. that
\begin{equation}
\label{eq:Hcar}
H=\frac{1}{2} \left(p_x^2+p_y^2\right)+V_1(x)+V_2(y)
\end{equation}
is the most general Hamiltonian function  in $T^*E_2$ that separates in Cartesian coordinates. In order to get the most general form of the integral of motion admitted by the Hamiltonian \eqref{eq:Hcar}, we solve the system  (\ref{cartA1}),  observing that
 \begin{equation}
\label{ }
k:=\frac{p_y^2}{2}+V_2(y),
\end{equation}
is a particular solution, functionally independent of the Hamiltonian \eqref{eq:Hcar}. Then, we can write down the general solution of the system
(\ref{cartA1})  as
\begin{eqnarray}
\nonumber f&=&\frac{x}{x-y}F(h,k) +\int F_{h} dk+G(h) \\
\nonumber g&=&-\frac{F }{x-y} \qquad x\neq y \ ,
\end{eqnarray}
where $h:= H$ and
%\begin{eqnarray}
% \nonumber h&:=&\frac{1}{2}\left( p_x^2+p_y^2\right) +V_1(x)+V_2(y),\\
% \nonumber k&:=&\frac{p_y^2}{2}+V_2(y),
%\end{eqnarray}
 $F=F\left(h, k \right)$,  $G=G \left(h \right)$ are arbitrary smooth functions of their arguments. The resulting expression for the integral of motion is given by a primitive function of the one--form (\ref{eq:alphacart})
\begin{equation} \label{Int2Cart}
H_2:= \int F ( h, k) dk + \int G (h) dh \ ,
\end{equation}
which is independent of $H_1:=H$ as
$$
\rd H_1 \wedge \rd H_2= F\,  \rd h \wedge \rd k =0 \quad  \Leftrightarrow g=0 \ .
$$
In particular, if $G=0$ and $F=1$ we get  the energy associated with the coordinate $y$
\begin{equation} \label{Int2CartSpe}
H_2=k=\frac{p_y^2}{2}+V_2(y).
\end{equation}
\begin{remark}
Let us observe that $g=1$, for any choice of $f$, does not solve Eqs. (\ref{cartA1}). Therefore,   neither a Lenard chain  nor a QBH chain generated by $(\omega, N_{car},H)$ can  exist. The same considerations apply to the polar and parabolic cases.
\end{remark}
\subsubsection{Polar case}

\noindent Let us consider the  natural Hamiltonian
\begin{equation} \label{eq:Hpol}
H=\frac{1}{2} \left(p_r^{2}+\frac{p_{\theta}^2}{r^2}\right)+V \left(r,\theta\right)\ ,
\end{equation}
where $(r,\theta,p_r,p_\theta)$ denote polar coordinates and their conjugate momenta.
A Nijenhuis operator that separates polar coordinates is
\begin{equation*}
N_{pol}=\mathrm{diag}(r,\theta,r,\theta),
\end{equation*}
with
\begin{equation} \label{eq:alphapol}
\alpha=f \left(r,\theta,p_r,p_{\theta}\right) dH+g\left(r,\theta,p_r,p_{\theta}\right) N^{T}dH.
\end{equation}
The closure condition for $\alpha$ provides the system (\ref{polA1}) reported in the Appendix.
By combining these equations
%(getting $V_r$ from the fourth one and $V_\theta$ from the second one, substituting in the last one, exploiting the first one) %
we get the consequence
\begin{equation} \label{Vpol}
g(\theta -r)\left(V_{r \theta }+\frac{2Ê}{r}V_\theta\right ) =0.
\end{equation}
The general solution of Eq. (\ref{Vpol}) ( $g\neq 0$, $r\neq \theta$) is
\begin{equation} \label{Vsol}
V(r, \theta )=V_1(r)+\frac{V_2(\theta) }{r^2 } \ ,
\end{equation}
which is the most general potential on $E_2$ that separates in polar coordinates.
As in the Cartesian case, we note that
\begin{equation}
\label{ }
k:= \frac{p_{\theta}^2}{2}+V_2(\theta)
\end{equation}
is a particular solution of the system (\ref{polA1}),  independent of the Hamiltonian \eqref{eq:Hpol}. Then we can state that
the general solution of the system (\ref{polA1}) is
\begin{eqnarray}
\nonumber f&=&\frac{r^3}{r-\theta} F(h,k) +\int F_{h} dk+G(h), \\
\nonumber g&=&-\frac{r^2}{r-\theta} F(h,k),
\end{eqnarray}
where $h:=H$
%\begin{eqnarray}
%\nonumber h:=&&\frac{1}{2} \left(p_r^{2}+\frac{p_{\theta}^2}{r^2}\right)+V_1(r)+\frac{V_2(\theta) }{r^2 }, \\
%\nonumber k:= &&\frac{p_{\theta}^2}{2}+V_2(\theta).
%\end{eqnarray}
and  $F=F\left(h, k \right)$, $G=G\left(h \right)$ are arbitrary smooth functions of their arguments. Therefore, the most  general expression of the integral  of motion reads
\begin{equation} \label{Int2Pol}
H_2:= \int F(h,k) dk +\int G(h) dh ,
\end{equation}
which is independent of $H_1:=H$ as
$$
\rd H_1 \wedge \rd H_2= F\,  \rd h \wedge \rd k =0 \quad  \Leftrightarrow g=0 \ .
$$
As a particular case, if $G=0$ and $F=1$ we obtain the simple expression of a generalized area integral
\begin{equation} \label{Int2PolSpe}
H_2=k=\frac{p_{\theta}^2}{2}+V_2(\theta).
\end{equation}
%\begin{remark}
%If $V_1= -c/r$ and $V_2=0$, $G=0$, $F=1$ we get a GL chain for the Kepler system, generated by $(\omega, N_{pol}, H)$. Other bi-Hamiltonian formulations  proposed in \cite{MarmoV, BW} do not produce any information about separation of variables.
%\end{remark}
%We will return to the Kepler problem at the end of Section VI.

\subsubsection{Parabolic case}

\noindent We will study the Hamiltonian
\begin{equation}
H=\frac{1}{2} \ \frac{p_\xi^{2}+p_{\eta}^2}{\xi^2+\eta^2 }+V \left(\xi,\eta\right),
\end{equation}
where $\xi , \eta$ are parabolic coordinates given by
\[
x= \frac{1}{2}(\xi^2-\eta^2), \qquad y=\xi \eta, \quad \xi \in \mathbb{R}, \quad \eta \geq 0,
\]
and $p_\xi , p_\eta$ their conjugate momenta.
Let us take the  Nijenhuis tensor
\begin{equation} \label{eq:Npar}
N_{par}=\mathrm{diag}(\xi, \eta,\xi,\eta),
\end{equation}
with
\begin{equation} \label{eq:alphapar}
\alpha=f \left(\xi , \eta , p_\xi,p_{\eta}\right) dH+g\left(\xi , \eta , p_\xi,p_{\eta}\right) N^{T}dH.
\end{equation}
The closure condition for $\alpha$ provides the system (\ref{parA1}) of the Appendix.
By combining  Eqs.  (\ref{parA1}) we get the consequence
\begin{equation} \label{Vpar}
g(\eta-\xi) \left( V_{\xi \eta }+2 \frac{Ê\eta V_\xi+\xi V_\eta }{\xi^2+\eta^2}\right )=0.
\end{equation}
The general solution of (\ref{Vpar}) ( $g\neq 0$, $\xi\neq \eta$)  is
\begin{equation} \label{Vparsol}
 V(\xi , \eta )=\frac{V_1(\xi)+V_2(\eta) }{\xi^2+\eta^2}
\end{equation}
which is the most general potential on $E_2$ that separates in parabolic coordinates. By following the same procedure as in the previous two cases, we deduce  the general solution of  system  (\ref{parA1})
%Taking into account (\ref{Vparsol}), system (\ref{parA1}) can be trasformed by means of a nonlinear transformation
%\begin{equation*}
%(\xi, \eta ; p_\xi , p_\eta )
%\mapsto (\xi, \eta ; h , k)
%\end{equation*}
%where
%\begin{eqnarray}
%h:=\frac{1}{\xi^2+\eta^2 } \left(\frac{p_\xi^{2}+p_{\eta}^2}{2}+ V_1(\xi)+V_2(\eta) \right).
%\\
% k:=\frac{1}{{\xi^2+\eta^2 }} \left( \frac{\eta^2 p_\xi^{2}-\xi^2 p_{\eta}^2}{2 } +
%\eta^2 V_1(\xi)-{\xi}^2 V_2(\eta)\right)
%\end{eqnarray}
%We get the equations (\ref{parA2}). The general solution of this system provides

\begin{eqnarray}
\nonumber f&=&-\frac{\xi^3 +\eta^3}{\xi-\eta} F(h,k) +\int F_{h} dk+G(h) \\
\nonumber g&=&\frac{\xi^2 +\eta^2}{\xi-\eta} F(h,k) \qquad \xi \neq \eta
\end{eqnarray}
where
\begin{eqnarray}
\nonumber h:=\frac{1}{\xi^2+\eta^2 } \left(\frac{p_\xi^{2}+p_{\eta}^2}{2}+ V_1(\xi)+V_2(\eta) \right)
\\
\nonumber k:=\frac{1}{{\xi^2+\eta^2 }} \left( \frac{\eta^2 p_\xi^{2}-\xi^2 p_{\eta}^2}{2 } +
\eta^2 V_1(\xi)-{\xi}^2 V_2(\eta)\right)
\end{eqnarray}
with $F=F\left(h, k \right)$ and $G=G\left(h \right)$ arbitrary smooth functions of their arguments. Consequently,
the most general  integral of motion admitted by a potential separable in parabolic coordinates can be represented as

\begin{equation} \label{Int2Par}
H_2:= \int F(h,k) dk +\int G(h) dh
\end{equation}
which is independent on $H_1:=H$ as
$$
\rd H_1 \wedge \rd H_2= F\,  \rd h \wedge \rd k =0 \quad  \Leftrightarrow g=0 \ .
$$

\noindent For $G=0$ and $F=1$ it follows that
\begin{equation} \label{Int2ParSpe}
H_2=k.
\end{equation}

\section{Integrable cubic H\'enon--Heiles systems}

In this section, we discuss the bi--Hamiltonian geometry of the integrable   cubic
H\'enon--Heiles systems.  We will show that they are conveniently described by the previous theory, and that they admit non trivial bi--Hamiltonian structures. The approach we follow is not, like the above, an inverse one (i.e. from separation coordinates toward the Hamiltonian system). Instead, we assume the explicit form of the Hamiltonians and construct the Nijenhuis
tensor fields and the corresponding chains together with the separation variables.

For a nice review on integrable H\'enon--Heiles systems, see \cite{CMV} and references therein. Besides, new integrable perturbations of these systems have been recently obtained in \cite{BB} by means of a Poisson algebra--type approach.

\par
The family of cubic H\'enon--Heiles systems is defined by the Hamiltonian function
\beq  \label{eq:HH}
H:=\frac{1}{2}\left(p_x^2+p_y^2\right)+\frac{1}{2}(c_1 x^2 +c_2 y^2) +a x y^2 -\frac{1}{3} b x^3  \ ,
\eeq
which is known to be integrable only in three cases:
%\begin{itemize}
% \item[i)]
%  $b=-a$  ;
%  \item [ii)]
%  $b=-6a$ ;
%  \item [iii)]
%  $b=-16a$.
%\end{itemize}
\begin{eqnarray}
\textit{(SK)}\qquad b & = & -a \quad\quad c_1=c_2=c \label{eq:SK}\\
\textit{(KdV)} \qquad b & = & -6a\quad \quad c_1, c_2 \ \text{arbitrary} \label{eq:KdV}\\
\textit{(KK)} \qquad b&=&-16 a  \quad\quad  c_1=16 c_2 \label{eq:KK}
\end{eqnarray}
They correspond, respectively, to stationary reduction of the fifth order flow of Sawada--Kotera (SK), Korteweg de Vries (KdV), Kaup--Kupershmidt (KK) soliton hierarchies \cite{Fo}.
In these tree cases an   integral of the motion is known
\begin{eqnarray}
H_2^{(SK)} & = & p_x p_y+\frac{1}{3} a \left(3 x^{2} y+y^{3}\right)+ c x y \label{eq:iSK}, \\
\nn  H_2^{(KdV)} & = &p_y(y p_x  -x p_y)+ \frac{1}{a}(c_2-\frac{c_1}{4})p_y^2+c_2 x y^2 +\\
&& +a(\frac{y^2}{4} +ax^2 )y^2 + \frac{c_2}{a}(c_2-\frac{c_1}{4})y^2 \label{eq:iKdV},\\
\nn H_2^{(KK)}&=&p_y^4 +p_y^2(2c_1 y^2+4axy^2)-\frac{4}{3}ap_x p_yy^3+\\
&-&\frac{4}{3}a^2 x^2y^4+c_1y^4-\frac{2}{9}a^2y^6 \label{eq:iKK}.
\end{eqnarray}

%\begin{itemize}
%  \item[i)]
%  $H_2=$ ;
%  \item [ii)]
%  $H_2=y p_xp_y -x p_y^2+\frac{a}{4} y^4+ax^2 y^2$;
%  \item [iii)]
%  $H_2=$.
%\end{itemize}

The SK cases (\ref{eq:SK}) and the KdV case (\ref{eq:KdV} have long been   known to be separable in rotated Cartesian coordinates \cite{AzSa} and parabolic coordinates \cite{W}, respectively. In contrast, only recently has  the KK case (\ref{eq:KK})  been proved to be separable in \cite{RGC}, by means of algebraic--geometric methods.

\subsection{SK--H\'enon--Heiles system}
In order to construct the Nijenhuis tensor field for the SK model  (\ref{eq:SK}), we shall discuss the general case of Hamiltonian systems separating by means of linear transformations of the plane. We have the following result.
\begin{proposition}
The most general Nijenhuis tensor field associated with systems separating in the
\textit{normal coordinates}
\beq
\chi_1= a_1 x +a_2 y, \quad \chi_2= a_3 x + a_4 y \ ,  \qquad a_1 a_4-a_2 a_3\neq 0
\eeq
with associated momenta
\begin{eqnarray}
\nn p_{\chi_1}=\frac{1}{a_1 a_4 - a_2 a_3}\left(a_4 p_x - a_3 p_y \right), \\
\quad p_{\chi_2}=\frac{1}{a_1 a_4 - a_2 a_3}\left(-a_2 p_x + a_1 p_y \right),
\end{eqnarray}
has the form
\begin{equation}
N_{norm}=\frac{1}{a_1 a_4-a_2 a_3}\left[\begin{array}{cccc}
n_{1,1} & n_{1,2}  & 0 & 0 \\
n_{2,1} & n_{2,2}  & 0 & 0 \\
0 & 0 & n_{1,1}   & n_{2,1}  \\
0 & 0 & n_{1,2}  & n_{2,2}
\end{array}\right] \ ,
\end{equation}
where
\bea
\noindent \nn n_{1,1} &=& (a_1^2 a_4 -a_2 a_3^2) x + a_2 a_4(a_1  -  a_3) y, \\
\nn n_{1,2}&=& a_2 a_4 [(a_1-a_3)x+(a_2-a_4)y)]  ,\\
\nn n_{2,1}&=& -a_1 a_3 \left[(a_1-a_3)x+(a_2-a_4) y \right], \\
\nn n_{2,2}&=&a_1 a_3( a_4- a_2 ) x +(a_1 a_4^2 -a_2^2 a_3) y  \ .\\
%\nn n_{3,3}&=&\nn n_{1,1} ,\\
%\nn n_{3,4}&=& \nn n_{2,1} ,  \\
%\nn n_{4,3}&=& \nn n_{1,2} , \\
%\nn n_{4,4}&=&\nn n_{2,2}
\eea
\end{proposition}

\noindent The case of the system \eqref{eq:SK} is obtained choosing $a_1=a_2=1/\sqrt{2}$ and $a_3=-a_4=1/\sqrt{2}$. The corresponding Nijenhuis tensor field is given by

\begin{equation} \label{eq:NSK}
N_{SK}=\frac{1}{\sqrt{2}} \left[\begin{array}{cccc}
 x &  y & 0 & 0  \\
 y &   x & 0 & 0 \\
0 & 0 &  x &   y \\
0 & 0 &  y &   x
\end{array}\right] \ '
\end{equation}
 the GL chain generated by $(\omega,N_{SK},H^{(SK)})$  is defined by
\beq
f=-\frac{x}{y},\qquad  g=\frac{\sqrt{2}}{y} \ , \quad y \neq 0,
\eeq
and  produces the integral of motion \eqref{eq:iSK}.

We recall that  in \cite{BW}, a different  bi--Hamiltonian structure  was proposed for the  SK case (\ref{eq:SK}). However,  being constant, it  gives no information about  separated variables.

\subsection{KdV--H\'enon--Heiles system}

In the KdV case \eqref{eq:KdV} we can put $c_1=c_2=0$ without loss of generality, $c_1\neq 0$, $c_2\neq 0$ corresponding to a shift of the parabolic web along the $y$--axis. By transforming the Hamiltonian function \eqref{eq:HH} and  \eqref{eq:KdV} in parabolic coordinates we find that it can be described by potential  of type \eqref{Vparsol} with
$$
V_1(\xi)=\frac{a}{4} \xi^8 \ , \qquad V_2(\eta)=-\frac{a}{4} \eta^8
$$
The GL chain generated by $\left(\omega, N_{par}, H^{(KdV)}_{|_{c_1=c_2=0}}\right)$ provides the integral \eqref{Int2ParSpe} that, in Cartesian coordinates, coincides with \eqref{eq:iKdV}.
\par
Let us observe that the Nijenhuis tensor field introduced for the first time in \cite{CRG}, \cite{BCRR}
in order to construct a QBH formulation of the KdV--Henon--Heiles case is nothing but
 \begin{equation} \label{eq:NKdV}
N_{KdV}=\mathrm{diag}(\xi^2,-\eta ^2 , \xi^2, -\eta ^2 ) \ .
\end{equation}
On the basis of Remark \ref{rem:st}, we can state that it defines the same parabolic web as  \eqref{eq:Npar}. However, this result does not contradict Remark 3, since the two Nijenhuis tensors \eqref{eq:Npar} and \eqref{eq:NKdV} depend on two different realizations of the same web.

\subsection{KK--H\'enon--Heiles system}

In order to construct a  Nijenhuis tensor field for the KK case (with $c_1=0$ and $a=1/4$), there are (at least) two possible procedures. The first one entails the use of a well--known canonical transformation between the SK and KK cases, that has been introduced and studied in \cite{RGC, BW, SEL}. As we have verified, this transformation directly maps the Nijenhuis structure of the SK case into that of the KK one.

However, we prefer to follow here  a different procedure. Precisely, we will consider a \textit{mixed problem}, in which the Hamiltonian of the KK case and its independent integral are
\begin{eqnarray}
\mathcal{H} ^{(KK)}& = & \frac{1}{2}(p_x^2 +p_y^2)+ \frac{1}{4}x  y^2+ \frac{4}{3}x^3, \\
\nn \mathcal{H} _2^{(KK)}& = & p_y^4 +p_y^2 x y^2- \frac{1}{3}p_x p_y y^3-\frac{1}{12} x^2 y^4 \\&-& \frac{1}{72} y^6,
\end{eqnarray}
and the Nijenhuis tensor field is to be determined. In our context, the geometrical relevance of the KK system is that it does not belong to the St\"ackel class, i.e. it is not separable in orthogonal coordinates in the plane: its first integral is a fourth--order one. Moreover, we observe that, in contrast to what occurs in all the  other examples of this paper,  the tensor \eqref{eq:NKK} is not a complete lift to the cotangent bundle (of the configuration space) of a  torsionless $L$--endomorphism defined on the configuration space \cite{IMM,BFP}.

Therefore, we must assume a general expression for the Nijenhuis tensor field in terms of a matrix $N_{KK}\in \mathcal{M}_{4\times 4}$, instead of a diagonal form.
Then, one has to impose three conditions: {\bf i)} that $N_{KK}$ be compatible with the canonical symplectic structure $\omega$; {\bf ii)} that $N_{KK}$ be torsionless; and {\bf  iii) } that $N_{KK}$ be compatible with the GL chain (partially determined by $\mathcal{H} ^{(KK)}$ and $\mathcal{H} _2^{(KK)}$) we wish to construct. The algebraic part of the condition {\bf i)}  imposes to the Nijenhuis tensor the form

\begin{equation} \label{eq:NKK}
N_{KK}= \begin{bmatrix}
     n_{1,1} & n_{1,2}&0&n_{1,4} \\
     n_{2,1}& n_{2,2}&-n_{1,4} &0 \\
      0&n_{3,2}&n_{1,1}&n_{2,1} \\
    -n_{3,2} & 0 & n_{1,2} & n_{2,2} \\
      \end{bmatrix} \ ,
\end{equation}
where $n_{i,j}$ are arbitrary  functions on $T^*E_2$.
The remaining  requirements altogether determine a complicated system of 20 nonlinear  partial differential equations plus four algebraic equations  (not reported here).
In order to find a particular solution of the determining system, a natural ansatz is to suppose that the entries of $N_{KK}$ be rational functions of the phase space variables.

By solving the systems of the 20 differential equations [corresponding to  conditions {\bf i)} and {\bf ii)}] we get the following structure for the Nijenhuis tensor:
\bea
\nn n_{1,1} &=&x, \quad n_{1,2}=\frac{k_1}{\phi(y)}+k_0 p_x p_y \phi'(y) \ ,\\
\nn n_{2,1}&=& \left(k_0 k_{2}^2 x-\frac{3 x^2}{k_1}+2 \frac{x k_3}{k_1}+k_4\right)\phi(y)\ , \\
\nn &-&\phi(y) \int\frac{d x_2}{\phi(y)} \ ,\\
\nn n_{2,2}&=&-2 x+k_3-k_0 k_2 p_y \phi(y)-\frac{k_0 p_y^{2}\phi(y)^2}{2 k_1}\ , \\
\nn n_{1,4}&=& k_0 p_x \phi(y)\ ,\\
\nn n_{3,2}&=&p_y\left[1+\left(k_0 k_2^{2} x - \frac{3 x^{2}}{k_1}+
\frac{2 x k_3}{k_1}+k_4 - \int\frac{d y}{\phi(y)}\right)\phi'(y)\right] \\
 &+&\frac{k_1 k_2 }{\phi(y)} \ ,
\eea
where $\phi(y)$ is an arbitrary  function and $k_0, \ldots , k_4$ are  arbitrary real constants.The four algebraic determining equations [condition {\bf iii)}], which are  specific for the chain of the KK model, restrict the previous solution and yield

%From Eq.\eqref{1} and \eqref{2} we deduce algebraically the (cumbersome) expressions of $f$ and $g$ in terms of the only unknown function $\phi(x_2)$. Once we replace these (very long) expressions into eqs. %\eqref{3} and \eqref{4}, both equations give exactly the same condition on $\phi(x_2)$. By annihilating the coefficient of the various monomials in the powers of $p_1,p_2,x_1$, we obtain immediately
\beq
\nn \phi(y)=\frac{3}{y},
\eeq
with
\beq
\nn k_0=1, \quad k_1=\frac{3}{4},\quad k_2=0,\quad k_3=0, \quad  k_4=0.
\eeq
Then we get the following structure for the GL chain of the KK model
\beq
f=-\frac{2}{3}\left(x y^{2}+6 p_y^{2}\right),
\eeq
\beq
g=-\frac{4}{3}y^{2}.
\eeq

%\begin{equation}
%f=-\frac{x y^2+6p_y^2} { \mathcal{H} _2^{(KK)}} \ , \quad g=- \sqrt{2} \frac{y^2}{ \mathcal{H} _2^{(KK)}} \ ,\label{GLKK}
%\end{equation}
%\noindent whereas for the Nijenhuis tensor we obtain
%\bea
%\noindent \nn n_{1,1} &=& \sqrt{2} x, \\
%\nn n_{1,2}&=&  \frac{\sqrt{2}}{Ê4}  (y-12 \frac{p_y p_x}{y^2}),\\
%\nn n_{1,4}&=&3\sqrt{2}\frac{p_x}{y}, \\
%\nn n_{2,1}&=&-\frac{\sqrt{2}}{2}(24\frac{ x^2}{y}+y), \\
%\nn n_{2,2}&=& -2\sqrt{2}(x+3\frac{p_y^2}{y^2} ),\\
%\nn n_{3,2}&=& \frac{3}{\sqrt{2}}(8\frac{x^2}{y^2}+1)p_y.
%\eea
\noindent The eigenvalues of \eqref{eq:NKK}
$$
\lambda_{1,2}= -\frac{1}{2}\left(x+6\frac{p_y^2}{y^2}\right )\mp\frac{3}{y^2}\sqrt{\mathcal{H} _2^{(KK)}},
$$
together with the conjugate momenta
$$
\mu_{1,2}=
\left( -p_{{x}}+6\,{\frac {p_{{y}}x}{y}}+12
\,{\frac {{p_{{y}}}^{3}}{{y}^{3}}}\pm \, 12 \frac {p_y}{Êy^3}
 \sqrt {\mathcal{H} _2^{(KK)}} \right),
$$
are separated coordinates for the KK--H\'enon--Heiles system. In fact, they allow to write down the following Sklyanin separation equations \eqref{eq:Sk} to be written

\begin{eqnarray}
\mu_{1}^2 &=&  -\frac{8}{3} \lambda_1^3+2\,\mathcal{H}^{(KK)}+ \sqrt {\mathcal{H} _2^{(KK)}}, \\
\mu_{2}^2 &=&   -\frac{8}{3} \lambda_2^3+2\,\mathcal{H}^{(KK)}- \sqrt {\mathcal{H}_2^{(KK)}}  \ .
\end{eqnarray}
\par
We recall that the above separated coordinates  coincide with those introduced in \cite{RGC, SEL} and the  Nijenhuis tensor field \eqref{eq:NKK} coincide with a particular case of the one  introduced in \cite{GT},  where it has been obtained by  a completely different method.

\section{Multi--Separation of Variables and superintegrable systems}

In this section, we use the bi--Hamiltonian structures constructed in the previous discussion to construct potentials admitting more than a system of separation coordinates.  In fact, it can achieve that a Hamiltonian function belongs to GL chains generated by different  and incompatible bi--Hamiltonian structures. In this case, we get a Hamiltonian system separable  in different coordinate system  or a \emph{multi--separable} system together with additional integrals of motion that, if they are independent, assures superintegrability of the model.  Thus, we recover in a natural way the Smorodinsky--Winternitz potentials in the plane, first discovered in a quantum--mechanical context in \cite{FMSUW}, \cite{Winternitz} and studied again in \cite{STW} and \cite{TTW} from a group theoretical point of view. These are the only potentials that are multi--separable in terms of orthogonal coordinates in $E_{2}$.

\subsection{Cartesian and Polar coordinates}

Let us search for the most general potential $V(x,y)$ that admits SoV both in Cartesian and in polar coordinates. To this end, let us write down  Eq. (\ref{Vpol}) (with $g\neq 0$) in Cartesian coordinates. It reads
\begin{equation}
\frac{1}{\sqrt{x^2+y^2}}\left(xy(V_{xx}-V_{yy })-(x^2-y^2)V_{xy}+3yV_x-3xV_y\right)=0.
\end{equation}
Thus, the potential  has to satisfy the system of two PDEs
\begin{equation}
V_{xy}=0,\\ \label{Carpol1}
\end{equation}
\begin{equation}
xy(V_{xx}-V_{yy })-(x^2-y^2)V_{xy}+3yV_x-3xV_y=0. \label{Carpol2}
\end{equation}
By substituting the solution of the Eq. \eqref{Carpol1}
\begin{equation}
V(x,y)=V_1(x)+V_2(y)
\end{equation}
into Eq. \eqref{Carpol2} we get the separated equations
\begin{equation}
V_1''+\frac{3}{x}V_1'=V_2''+\frac{3}{y}V_2'=4a,
\end{equation}
where $a$ is an arbitrary constant. Their general solution is
\begin{eqnarray}
V_1(x)=\frac{1}{2}a x^2+\frac{ c_1}{x^2},\\
V_2(y)=\frac{1}{2}a y^2+\frac{ c_2}{y^2}.
\end{eqnarray}
Thus, the general solution of the system \eqref{Carpol1}--\eqref{Carpol2} is
\begin{equation} \label{VCarPol}
V(x,y)=\frac{1}{2}a( x^2+y^2)+\frac{ c_1}{x^2}+\frac{ c_2}{y^2},
\end{equation}
which is nothing but the SWI potential \cite{Winternitz}, sum of an isotropic elastic potential and an anisotropic Rosochatius potential. The Hamiltonian system with the SWI potential inherits the integral of motion (\ref{Int2CartSpe}) from SoV in Cartesian coordinates
\begin{equation}
H_2^{(car)=}\frac{p_y^2}{2}+\frac{a}{2}y^2+\frac{c_2}{y^2},
\end{equation}
and the integral (\ref{Int2PolSpe}) which, written in Cartesian coordinates reads
\begin{equation}
H_2^{(pol)}=\frac{1}{2}(xp_y-yp_x)^2+c_1\left( \frac{y}{x}\right)^2+c_2\left( \frac{x}{y}\right)^2.
\end{equation}
A simple check shows that the Hamiltonian SWI, $H_2^{(car)}$ and $H_2^{(pol)}$ are independent. Consequently, the potential (\ref{VCarPol}) is superintegrable.
Finally, we can state that the SWI Hamiltonian function  generates two GL chains, starting from the two incompatible  structures $(\omega , N_{car})$ and $(\omega , N_{pol})$. Indeed, it can be checked that the two Poisson tensors fields
$$
P_{car}:=N_{car} (\omega^\flat)^{-1} \ , \quad P_{pol}:=N_{pol} (\omega^\flat)^{-1}
$$
have non vanishing Schouten brackets, i.e.,
$$
\left [P_{car}, P_{pol} \right ]_S\neq 0  \ .
$$

\subsection{Cartesian and Parabolic coordinates}

By repeating the previous strategy, we find that the most general potential $V(x,y)$ that admits SoV both in Cartesian and in parabolic coordinates reads
\begin{equation} \label{VcartPar}
V(x,y)=a(4x^2+y^2)+c_1x +\frac{c_2}{y^2},
\end{equation}
which is nothing but the SWII potential \cite{Winternitz}.

%Alternatively, one can  write down eq. (\ref{Vpar}) (with $g\neq0$)  in Cartesian  coordinates. %It reads,
%\begin{equation}
%y(V_{yy}-V_{xx})+2x V_{xy}+3V_y=0.
%\end{equation}
%Thus, the potential  has to satisfy the system of two PDE
%\begin{eqnarray}\label{PolCart}
%V_{xy}=0 \\
%y(V_{yy}-V_{xx})+2x V_{xy}+3V_y=0
%\end{eqnarray}
%Substituting the solution of the first equation

%V(x,y):=V_1(x)+V_2(y)
%\end{equation}
%into the second equation we get the separated equations
%\begin{equation}
%V_1''=V_2''+\frac{3}{y}V_2'=8a
%\end{equation}
%where $a$ is an arbitrary constant. Its general solution is
%\begin{eqnarray}
%V_1(x)=4a x^2+c_1x\\
%V_2(y)=a y^2+\frac{ c_2}{y^2}
%\end{eqnarray}
%Thus, the general solution of the system (\ref{PolCar}) is (\ref{VcartPar}).

The Hamiltonian system with the SWII potential inherits the integral of motion (\ref{Int2CartSpe}) from SoV in Cartesian coordinates, 
\begin{equation}
H_2^{(car)}= \frac{p_y^2}{2} +\frac{a}{2} y^2+\frac{c_2}{y^2}
\end{equation}
and the integral (\ref{Int2Par}) from SoV in parabolic coordinates which, written in Cartesian coordinates,  reads
\begin{equation}
H_2^{(pol)}=p_y(yp_x-xp_y)+2axy^2+\frac{c_1}{2}y^2-2c_2\frac{x}{y^2}.
\end{equation}
By checking the functional independence of the Hamiltonian SWII,  $H_2^{(car)}$ and $H_2^{(par)}$, we obtain again that the potential (\ref{VcartPar}) is superintegrable. Moreover,
as in the  previous case, we can state that the SWII Hamiltonian function  generates two GL chains, starting from the two incompatible  structures $(\omega , N_{car})$ and $(\omega , N_{par})$. Indeed, it can be checked that the  Poisson tensors field
$$
 P_{par}:=N_{pol} (\omega^\flat)^{-1}
$$
has non vanishing Schouten brackets with $P_{car}$
$$
\left [P_{car}, P_{par} \right ]_S\neq 0  \ .
$$

In \cite{Ranada} a bi--Hamiltonian formulation for SWI and SWII has been proposed. The main difference with respect to our approach is that, since the two Poisson tensor fields obtained in \cite{Ranada} are not compatible [as they
 do not satisfy \eqref{eq:PQSch}], therefore, they do not define a Nijenhuis tensor field, and do not provide informations about the separated variables admitted by the systems.

\subsection{Polar and Parabolic coordinates}

The most general potential $V(r, \theta)$ that admits SoV both in polar and in parabolic coordinates, written down in polar coordinates, reads
\begin{equation} \label{VPolPar}
V(r,\theta)=\frac{\alpha}{r}+\frac{ 1}{ r^2}\frac{\beta+\gamma cos\theta}{ sin^2\theta},
\end{equation}
where
$$
\alpha=\frac{c_2}{2} \qquad
\beta:= \frac{1}{4}\left( c_3-\frac{c_1}{2}\right)\qquad
\gamma:= \frac{1}{4}\left( c_3+\frac{c_1}{2}\right)
$$
that is the SWIII potential \cite{Winternitz}.
%Alternatively, one can write down equation (\ref{Vpar}) (with $g\neq0$)  in Cartesian  coordinates.
%It reads,
%\begin{equation}
%V_\theta cos\theta-V_{\theta \theta}sin\theta+2rcos\theta V_{r\theta}+r^2sin\theta V_{rr }+V_r 2r sin\theta=0.
%\end{equation}
%Thus, the potential  has to satisfy the system of two PDE
%\begin{eqnarray*}\label{PolCart}
%\noindent && V_{r \theta }+\frac{2Ê}{r}V_\theta=0, \\
%\noindent && V_\theta cos \theta-V_{\theta \theta}sin\theta+2rcos\theta V_{r\theta}+r^2sin%%\theta V_{rr }\\
%& +V_r 2r sin\theta=0
%\end{eqnarray*}
%Substituting the solution of the first equation
%\begin{equation*}
%V(r,\theta):=V_1(r)+\frac{V_2(\theta)}{r^2}
%\end{equation*}
%into the second equation we get the separated equations
%\begin{equation}
%r^3\left( r V_1''+2V_1' \right)=V_2''+3cotg\theta V_2'-2V_2=a
%\end{equation}
%where $a$ is an arbitrary constant. Its general solution is
%\begin{eqnarray}
%V_1(r)=\frac{a}{2r^2}+\frac{\alpha}{ r},\\
%V_2(\theta)=\frac{\beta -a}{sin^2\theta}+\frac{\gamma cos\theta }{ sin^2\theta}-a \frac{cos2\theta+3}{2cos2\theta-2}.
%\end{eqnarray}
%Thus, the general solution of the system (\ref{PolParr}) is (\ref{VPolPar}).

The Hamiltonian system with the SWIII potential inherits the integral of motion (\ref{Int2PolSpe}) from SoV in polar coordinates
\begin{equation}
H_2^{(pol)}= \frac{p_\theta^2}{2} +\frac{\beta+\gamma cos\theta}{sin^2\theta} \label{H2pol}
\end{equation}
and the integral (\ref{Int2ParSpe}) from SoV in parabolic coordinates, which, written down in polar coordinates, reads
\begin{eqnarray}
\nonumber \noindent && H_2^{(par)}=-p_\theta \left(\frac{p_\theta cos\theta}{r} +p_r sin\theta \right )-\alpha cos\theta+ \\
\noindent && 2 \frac{\gamma+2\beta cos\theta+\gamma cos^2\theta}{ rsin^2\theta}
\end{eqnarray}
As in the previous cases, the potential (\ref{H2pol}) is superintegrable. Furthermore, also  the SWIII Hamiltonian function  generates two GL chains, starting from the two incompatible  structures $(\omega , N_{pol})$ and $(\omega , N_{par})$. Indeed, it can be checked that
$$
\left [P_{pol}, P_{par} \right ]_S\neq 0  \ .
$$

\subsubsection{The Kepler model}
The best known example of a Hamiltonian system separable both in polar and in parabolic coordinates is the  Kepler model
$$
H_K:=\frac{1}{2}\left(p_x^2+p_y^2\right)-\frac{a}{\sqrt{x^2
+y^2}}.
$$
Here, we recover from the theory illustrated above  a new bi--Hamiltonian structure and a
GL chain for this model. Precisely, by exploiting the freedom in the choice of separated coordinates discussed in Remark \ref{rem:st}, we take as a Nijenhuis tensor field that separates polar coordinates the following one,
$$
N_{polK}=\mathrm{diag}(r,\tan \theta, r, \tan \theta),
$$
that in Cartesian coordinates reads
\begin{equation}
N_{polK}=\left[\begin{array}{cccc}
n_{1,1} & n_{1,2}  & 0 & 0 \\
n_{1,2}& n_{2,2}  & 0 & 0 \\
0 &  n_{3,2}    & n_{1,1}  & n_{1,2} \\
-n_{3,2}   & 0 & n_{1,2}  &n_{2,2}
\end{array}\right] \ ,
\end{equation}
where
\bea
\noindent \nn n_{1,1} &=&\frac{y^3}{x r^2} + \frac{x^2}{r} \\
\nn n_{1,2}&=& -\frac{y^2}{r^2}+ \frac{ x y}{r }  ,\\
%\nn n_{2,1}&=& n_{1,2}, \\
\nn n_{2,2}&=&\frac{x y}{ r^2} + \frac{y^2}{r} ,\\
\nn n_{3,2}&=&(\frac{Êy^2}{x r^2Ê} -\frac{y}{r})p_x + ( \frac{-y}{r^2}+\frac{x}{r})p_y \ , \\
%\nn n_{3,3}&=& n_{1,1} ,\\
%\nn n_{3,4}&=& n_{2,1} ,  \\
%\nn n_{4,1}&=& -n_{3,2} , \\
%\nn n_{4,4}&=&n_{2,2}
\eea
\par
\noindent with $r=\sqrt{x^2+y^2}$. The GL chain generated by $(\omega, N_{polK}, H_K)$ with
$$
f=-\frac{x}{y-xr} r \ , \qquad\qquad g=\frac{x}{y-xr} r^2
$$
provides as a second integral of motion the square of the modulus of the angular momentum:
$$
H_2^{(polK)}=\frac{1}{2}(xp_y-yp_x)^2 \ .
$$
Furthermore, let us take as a Nijenhuis tensor field that generates the  parabolic web the following tensor,
\begin{equation} \label{eq:NparK}
N_{parK}=diag(\xi^2,-\eta ^2 , \xi^2, -\eta ^2 ) \ ,
\end{equation}
which in Cartesian coordinates has the linear representation
\begin{equation}
N_{parK}=\left[\begin{array}{cccc}
2x & y & 0 & 0 \\
y & 0 & 0 & 0 \\
0 & p_y   & 2x  & y  \\
-p_y & 0 &y  & 0
\end{array}\right] \ .
\end{equation}
\noindent The GL chain generated by $(\omega, N_{parK}, H_K)$ with
\[
f=2x\ , \qquad\qquad g=-1 \ ,
\]
provides as a second integral of motion the $x$--component of the Laplace--Runge--Lenz vector
\[
H_2^{(parK)}=p_y(xp_y-yp_x)-\frac{a}{\sqrt{x^2 +y^2}} x\ .
\]

Surprisingly, the Kepler Nijenhuis tensor  \eqref{eq:NparK}  coincides with the Nijenhuis tensor \eqref{eq:NKdV}, constructed  by other authors for the KdV--Henon--Heiles system.  This is due to the fact that $N_{parK}$ generates the same parabolic web in which both systems are separable.

In \cite{MV} a different bi--Hamiltonian formulation for the three--dimensional Kepler problem has been introduced. The main motivation of the authors was to prove that the existence of a recursion operator does not necessarily provide  additional conservation laws. The structures obtained in \cite{MV} are expressed in terms of action--angle variables, and give no information about the separated variables.

In \cite{SV} another recursion operator for the Kepler potential has been proposed. However, the difference with respect to our approach is that their recursion operator is not compatible with the canonical symplectic structure.

\section{Future perspectives}

In this work, we have proposed a general formalism for treating the geometry of both integrable and superintegrable systems on a bi--Hamiltonian setting. The present approach for the sake of concreteness has been formulated in the Euclidean plane. However, there is no theoretical restriction in extending it to higher--dimensional cases. Also, it seems interesting to include in the present analysis integrable and superintegrable systems defined in curved spaces.
It would be very interesting to derive a quantum formulation of the present theory.

 \section*{Appendix}
Here we report the explicit expressions of the systems of differential equations quoted in Sec. III.

a) Cartesian system.
\begin{eqnarray} \label{cartA1}
\nonumber\noindent &&  p_y (f_{p_x}+y g_{p_x})-p_x(f_{p_y}+x g_{p_y})=0, \\
\nonumber\noindent && p_x (f_{x}+x g_{x}+g)-(f_{p_x}+x g_{p_x})V_{x}=0, \\
\nonumber\noindent && p_y (f_{x}+y g_{x})-(f_{p_y}+x g_{p_y})V_{x}=0, \\
\nonumber\noindent &&  p_x (f_{y}+x g_{y})-(f_{p_x}+y g_{p_x})V_{y}=0, \\
\nonumber\noindent &&  p_y (f_{y}+y g_{y}+g)-(f_{p_y}+y g_{p_y})V_{y}=0, \\
\noindent && (x-y)g V_{x y}+(f_{y}+x g_{y})V_{x}-(f_{x}+y g_{x}) V_{y} =0 \text{.}
\end{eqnarray}
b) Polar system.
\begin{eqnarray}\label{polA1}
\nonumber \noindent &&\frac{p_{\theta}}{r^2} \left(f_{p_r}+\theta g_{p_r}\right)-p_r\left(f_{p_{\theta}}+r g_{p_{\theta}}\right)=0, \\
\nonumber\noindent &&  p_{r} \left(f_{\theta}+r g_{\theta}\right)-\left(f_{p_r}+\theta g_{p_{r}}\right)V_{\theta}=0, \\
\nonumber \noindent && \frac{p_{\theta}}{r^2} \left(f_{\theta}+\theta g_{\theta}+g\right)-\left(f_{p_{\theta}}+\theta g_{p_{\theta}}\right)V_{\theta}=0, \\
\nonumber \noindent && p_{r} \left(f_{r}+r g_{r}+g\right)+\frac{p_{\theta}^2}{r^3}\left(f_{p_r}+r g_{p_{r}}\right)-\left(f_{p_r}+ r g_{p_{r}}\right)V_{r}=0, \\
\nonumber \noindent && \frac{p_{\theta}}{r^2} \left(f_{r}+\theta g_{r} \right )+\frac{p_{\theta}}{r^3} 2 g\left( r-\theta \right) +\frac{p_{\theta}^2}{r^3} \left(f_{p_{\theta}}+r g_{p_{\theta}} \right) \\
\nonumber \noindent &&  -\left(f_{p_{\theta}}+r g_{p_{\theta}}\right ) V_{r}=0, \\
\nonumber\noindent && \left(\theta-r\right) g V_{r \theta}+\left(f_{r}+\theta g_{r} \right) V_{\theta}-\left(f_{\theta}+r g_{\theta}\right) V_{r} \\
\noindent && + \frac{p_{\theta}^2}{r^3}\left( f_{\theta}+ rg_{\theta} \right) =0\text{.}
\end{eqnarray}
c) Parabolic system.
\begin{eqnarray}\label{parA1}
\nonumber\noindent && p_\xi\left(f_{p_\eta } + \xi g_{p_\eta}\right) - p_\eta \left(f_{p_\xi } + \eta g_{p_\xi}\right) =0, \\
\nonumber\noindent &&  p_\xi \frac{ \left(f_\xi +\xi g_\xi +g \right) }{ \xi^2+\eta^2  }  +\left( \xi \frac {p_\xi^2+p_\eta^2} { \left(\xi^2+\eta^2 \right )^2 }  -
V_\xi  \right) \left(f_{p_\xi } + \xi g_{p_\xi}\right)    =0, \\
\nonumber\noindent &&   2 \xi p_\eta \frac{\left(f+\xi g \right) }{\left( \xi^2+\eta^2 \right)^2 }+p_\eta \left(\frac{ f+ \eta g}{\xi^2+\eta^2} \right )_\xi +\left( \xi \frac {p_\xi^2+p_\eta^2} {\left( \xi^2+\eta^2 \right)^2 }  -V_\xi  \right)
 \\
\nonumber\noindent && \times \left(f_{p_\eta } + \xi g_{p_\eta}\right)    =0, \\
\nonumber\noindent &&   2  \eta p_\xi\frac{\left(f+\eta g \right) }{\left( \xi^2+\eta^2 \right)^2 }+p_\xi \left(\frac{ f+\xi g}{\xi^2+\eta^2} \right )_\eta +
\left( \eta\frac {p_\xi^2+p_\eta^2} {\left( \xi^2+\eta^2 \right)^2 }  -V_\eta  \right)\\
\nonumber\noindent && \times \left(f_{p_\xi } + \eta g_{p_\xi}\right)    =0, \\
 \nonumber\noindent &&   p_\eta\frac{\left(f_\eta+\eta g_\eta +g\right) }{\left( \xi^2+\eta^2 \right) }+\left( \eta\frac {p_\xi^2+p_\eta^2} {\left( \xi^2+\eta^2 \right)^2 }  -V_\eta  \right) \left(f_{p_\eta } + \eta g_{p_\eta}\right)    =0, \\
\nonumber\noindent &&  g \left( \xi -  \eta  \right)V_{ \xi \eta}+ \left(f_\eta+\xi g_\eta \right)V_\xi- \left(f_\xi+\eta g_\xi \right) V_\eta+ \left(p_\xi^2 +p_\eta^2 \right )\\
\nonumber\noindent &&\times  \left[\eta\left(  \left( f+\eta g\right)\frac{p_\xi^2 +p_\eta^2  }{\left( \xi^2+\eta^2 \right)^2 }  \right)_\xi -\xi
\left( \left( f+\xi g\right)\frac{p_\xi^2 +p_\eta^2  }{\left( \xi^2+\eta^2 \right)^2 }  \right)_\eta \right] \\
\noindent && =0 \text{.}
\end{eqnarray}

{\bf ACKNOWLEDGMENTS}

The research of P. T. has been partially supported by the Grant No. FIS2011--22566, Ministerio de Ciencia e Innovaci\'on, Spain. The research of G. T. has been partly supported by the Grant PRIN--2008 of the Italian MIUR. G. T. also wishes  to thank the European Research Network MISGAM for financial support.

\end{document}